\documentclass{article} 
\usepackage{iclr2023_conference,times}

\usepackage{amsmath,amsfonts,bm}

\def\eqref#1{equation~\ref{#1}}

\def\1{\bm{1}}

\def\eps{{\epsilon}}

\DeclareMathAlphabet{\mathsfit}{\encodingdefault}{\sfdefault}{m}{sl}
\SetMathAlphabet{\mathsfit}{bold}{\encodingdefault}{\sfdefault}{bx}{n}

\usepackage{hyperref}
\usepackage{url}
\usepackage{lipsum}
\usepackage{graphicx} 
\title{Graph-informed simulation-based inference for models of active matter}

\author{Namid R. Stillman\\
Department of Cell and Developmental Biology\\
University College London\\
London, UK \\
\And
Silke Henkes\\  
Lorentz Institute for Theoretical Physics \\
Leiden University \\
Leiden, The Netherlands \\
\And
Roberto Mayor\\  
Department of Cell and Developmental Biology\\
University College London\\
London, UK \\
\And
Gilles Louppe\\ 
Montefiore Institute\\
University of Liège\\
Liège, Belgium\\
}

\iclrfinalcopy 
\begin{document}

\maketitle

\begin{abstract}
Many collective systems exist in nature far from equilibrium, ranging from cellular sheets up to flocks of birds. These systems reflect a form of active matter, whereby individual material components have internal energy. Under specific parameter regimes, these active systems undergo phase transitions whereby small fluctuations of single components can lead to global changes to the rheology of the system. Simulations and methods from statistical physics are typically used to understand and predict these phase transitions for real-world observations. In this work, we demonstrate that simulation-based inference can be used to robustly infer active matter parameters from system observations. Moreover, we demonstrate that a small number (from one to three) snapshots of the system can be used for parameter inference and that this graph-informed approach outperforms typical metrics such as the average velocity or mean square displacement of the system. Our work highlights that high-level system information is contained within the relational structure of a collective system and that this can be exploited to better couple models to data.
\end{abstract}

\section{Introduction}

The difference between animate and inanimate matter is fundamental to biological systems. However, many biological systems behave, over short time periods, in similar ways to classical materials. For example, confluent sheets of epithelial cells form a rigid substrate with quantifiable stiffness and murmurations of flocks of starlings can create swirling patterns that mimic the phase transition of liquid to gas. Collective systems of individual agents that  contain their own energy source, also known as active matter, are always far from equilibrium and cannot be understood using thermodynamic theory. Instead, a combination of computer simulation and statistical physics has typically been used to understand these systems and quantify their states. 

One of the hallmarks of active matter is the rise of emergent properties, system-wide features that exist across length-scales much larger than an individual entity. While the exact circumstances that lead to emergent properties is still an open question, there is some evidence that the connectivity of a system may be important. In this work, we build our own particle-based active matter model and use the system connectivity, represented as a contact graph, to infer the simulation parameters using simulation-based inference. We find that a single snapshot allows for robust inference, highlighting that high-level properties are contained within the structure of a collective system and that increasing the number of snapshots improves inference power. This work is the first, to the best of our knowledge, to combine geometric deep learning with simulation-based inference to quantify collective systems and demonstrates the strength of applying these methods when compared to conventional approaches. 

\subsection{Related Works}
The field of active matter has heavily relied upon computer simulation in order to identify phase transitions and system features over the last three to four decades. There are several comprehensive reviews which cover this topic, including \citet{ramaswamy2010mechanics}, \citet{shaebani2020computational} or \citet{das2020introduction}. In this work, we consider a 2D system that approximates the properties of collective cell migration. For other works that consider cellular systems as active matter, see, for example \citet{alert2020physical}, \citet{henkes2020dense} or \citet{stillman2023generative}.

The field of simulation-based inference can refer to both the use of simulation with Bayesian sampling to infer model parameters and the specific application of neural networks to approximate posterior densities from simulation. Whereas the former meaning reflects a large body of work, including approximate Bayesian computation and others (see, for example, \citet{marin2012approximate}), we refer specifically to the latter. For further details on simulation-based inference see, for example, the review given in \citet{cranmer2020frontier}. 

Finally, this work is, to the best of our knowledge, the first application of simulation-based inference to active matter. Previous work has used ABC to infer properties of cell migration, and other papers have used Bayesian inference to quantify model uncertainty \citep{vo2015quantifying, ross2017using}. However, none of these works have used recent advances in deep learning to improve the quality and statistical power of inference. With regards to graph-based learning we refer the reader to \citet{wu2020comprehensive} for an overview and note the recent work of \citet{dyer2022calibrating} on combining simulation-based inference with graph-based learning for agent-based models. 

\section{Methods}
\label{methods}

\subsection{Active Matter Model}
\label{sec:active_matter_model}
\begin{figure}[b]
	\begin{center}
	\includegraphics[width=\linewidth]{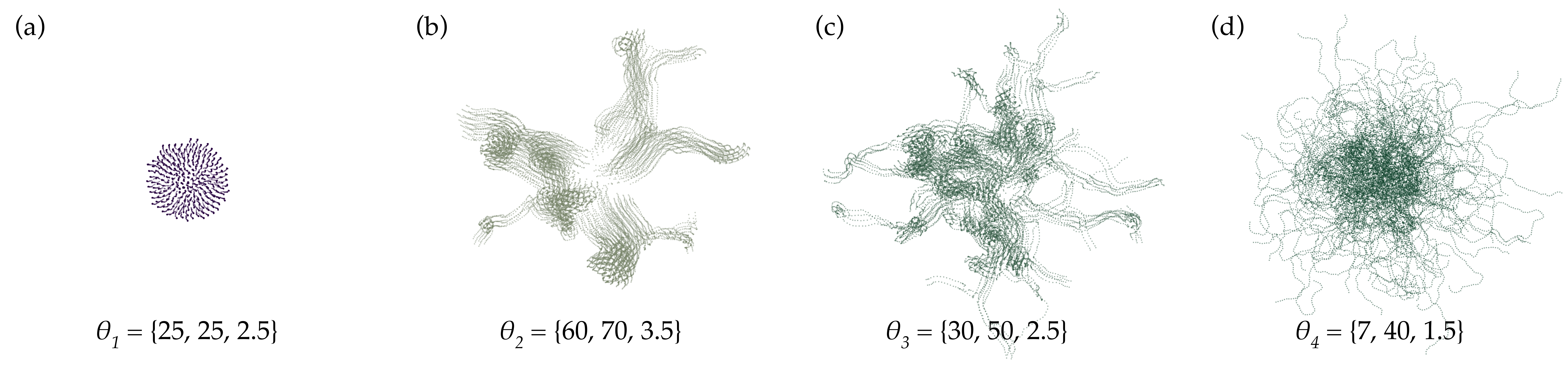}
	\end{center}
	\caption{Example outputs from simulations of active matter representing different phases, where we alter parameter values, $\theta$, for the active force, $v_0$, stiffness, $k$, and persistence timescale, $\tau$, of an active matter model. Hence, $\theta = \{v_0, k, \tau\}$.}
 \label{fig:active_matter_model}
\end{figure}

To test the applicability of simulation-based inference for active matter systems, we use a custom-built simulator to output trajectories of active Brownian particles, with details provided in \autoref{ap:model}. We assume that particles exert mechanical forces on neighbouring particles within some contact radius, according to an interaction potential that  contains both stiff and soft components. This interaction potential is also detailed in \autoref{ap:model}. Example output from our simulations can be seen in \autoref{fig:active_matter_model}. In this work, we consider parameter sets, $\theta$, which vary the active force, $v_0$, the stiffness of interactions, $k$, and the persistence timescale for a particle, $\tau$. 

\subsection{Simulation-based Inference}

We are interested in whether we can determine a set of parameters for an active matter system, given some observations. In Bayesian terms, this is equivalent to estimating the posterior probability distribution, $p(\theta| \bm{x})$, for parameters $\theta$ and observations $\bm{x}$. Unfortunately, most inference algorithms require the explicit calculation of the likelihood for the system, which is intractable for our case. Instead, we use amortized variational inference, or neural posterior estimation (NPE), to learn a neural density estimator of the posterior from training data generated from the joint distribution $p(\theta, \bm{x}) = p(\theta) p(\bm{x} | \theta)$. We use the LAMPE package to perform our simulation-based inference\footnote{Package details can be found here: https://lampe.readthedocs.io}, using a masked autoregressive flow network \citep{papamakarios2017masked} as density estimator.
Further details on the different architectures designs are given in \autoref{ap:architectures}. 
	
\subsection{Graph-informed Simulation-based Inference}

The output from the simulator includes position and velocity data of individual particles. We keep the number of particles fixed at 200 and output 100 snapshots of the system. Hence, output from the simulator is  relatively small in the context of active matter models but still must be summarised for simulation-based inference. Typical approaches for summarising active matter systems includes metrics such as the average velocity and the mean square displacement (MSD). In this work, we compare these classical approaches to summarising active matter with graph-based learning.

\begin{figure}[ht]
	\begin{center}
		\includegraphics[width=1\linewidth]{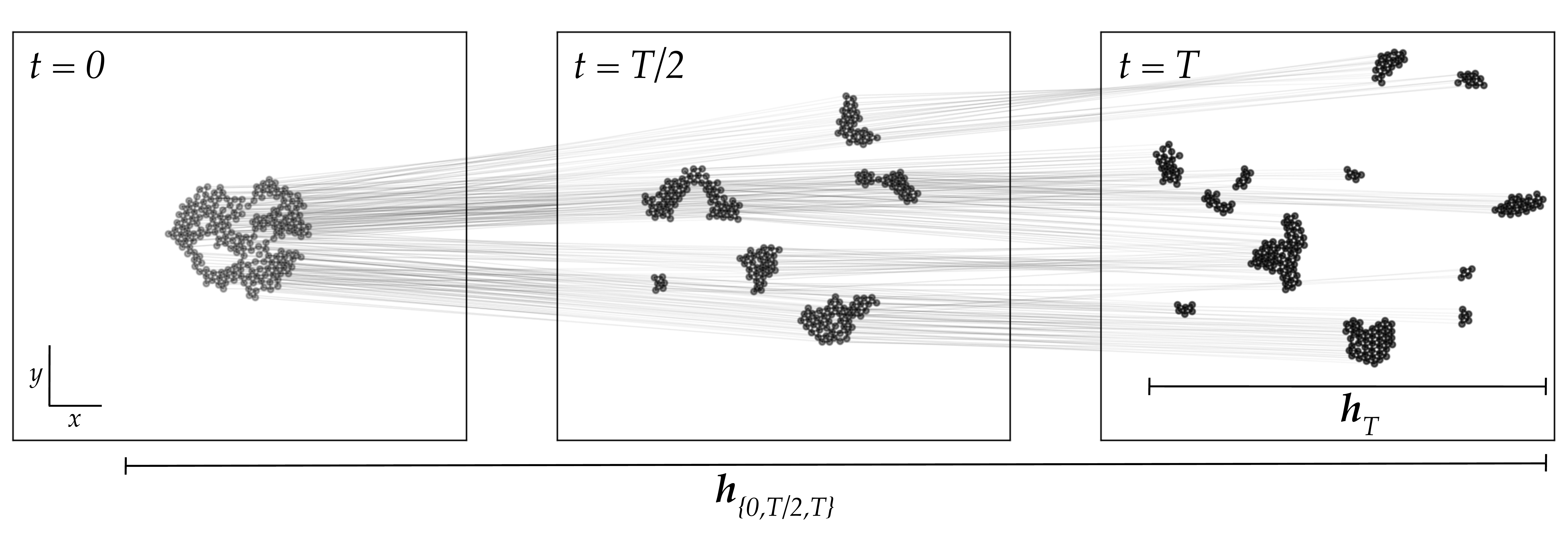}
	\end{center}
	\caption{We use spatio-temporal interaction networks to create our embeddings, where spatial interactions are based on a cutoff and where we connect the same particles between timesteps. Graph embeddings are constructed using either single snapshots, such as $\textbf{h}_T$ above, or all snapshots, as in $\textbf{h}_{\{0,T/2,T\}}$. Here, we show interaction network for simulation output in \autoref{fig:active_matter_model}(b).}
	\label{fig:2}
\end{figure}

Specifically, we assume that the state of the system for each timestep can be represented by a graph constructed from a set of nodes and edges, namely $\mathcal{G}_t = \{\mathcal{N}_t,\mathcal{E}_t\}$, where nodes are given by the particles spatial position and velocities and edges represent where two particles are within a specified contact radius, as described in  \autoref{sec:active_matter_model}. Example graphs for different states of the system are given in \autoref{fig:2}. For all nodes, we include position and velocity data. We use graph neural networks (GNNs) to embed specific timesteps of the system into low dimensional graph embedding, $\textbf{h}_t$, where we use graph convolutional networks \citep[GCNs,][]{kipf2016semi}, with global mean pooling. We consider two different cases, where we have only the final timestep, $\mathcal{G}_T$, and where we have the initial, final and a `halfway' timestep, $\mathcal{G}_{\{0,T/2,T\}}$. We then pass these graph embeddings to the neural network used for estimating the posterior. In practice, both embedding and density estimator networks are trained in parallel.

\section{Results}
\label{results}

\begin{figure}[t]
	\begin{center}
		\includegraphics[width=0.45\linewidth]{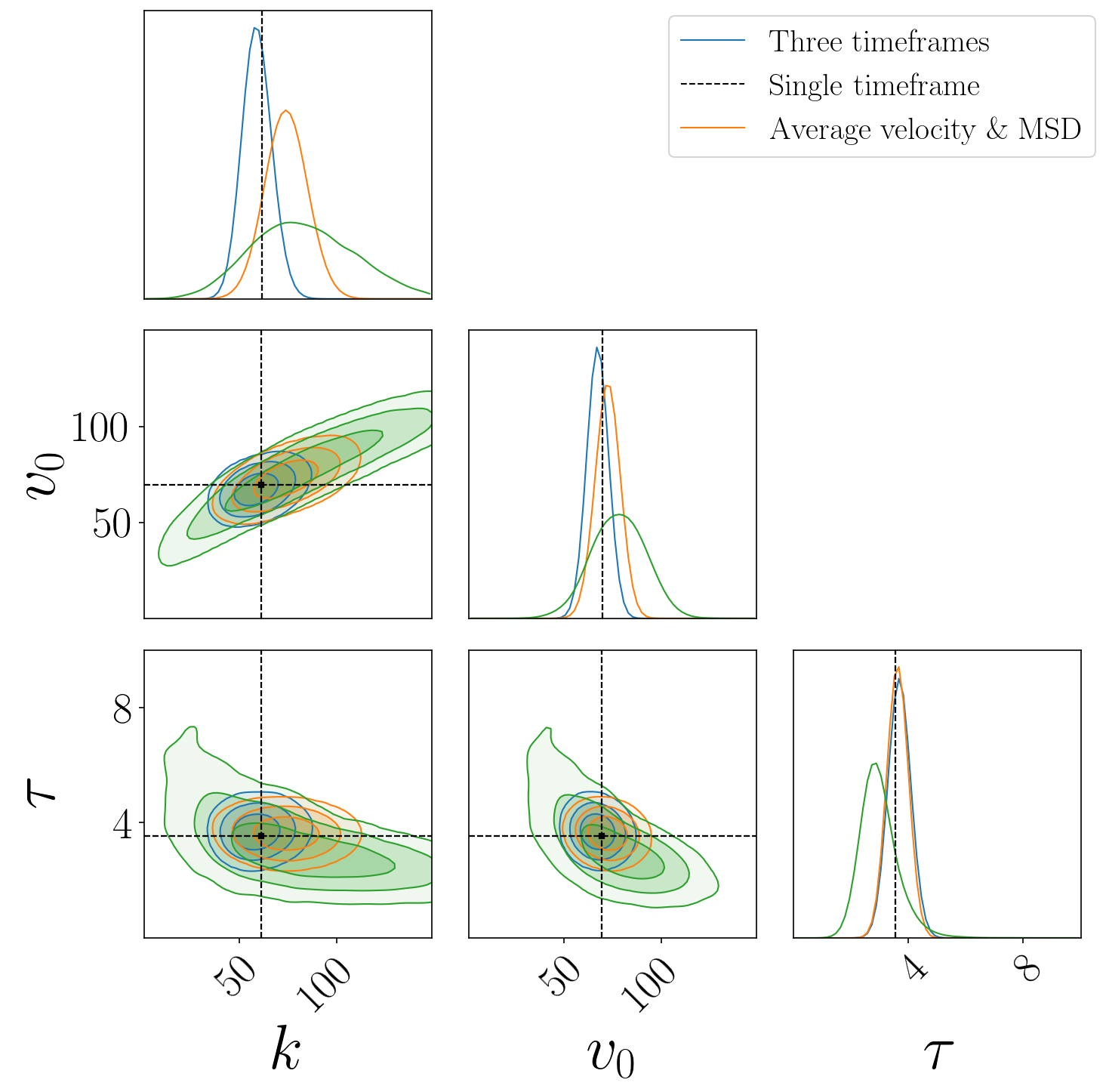}
		\includegraphics[width=0.45\linewidth]{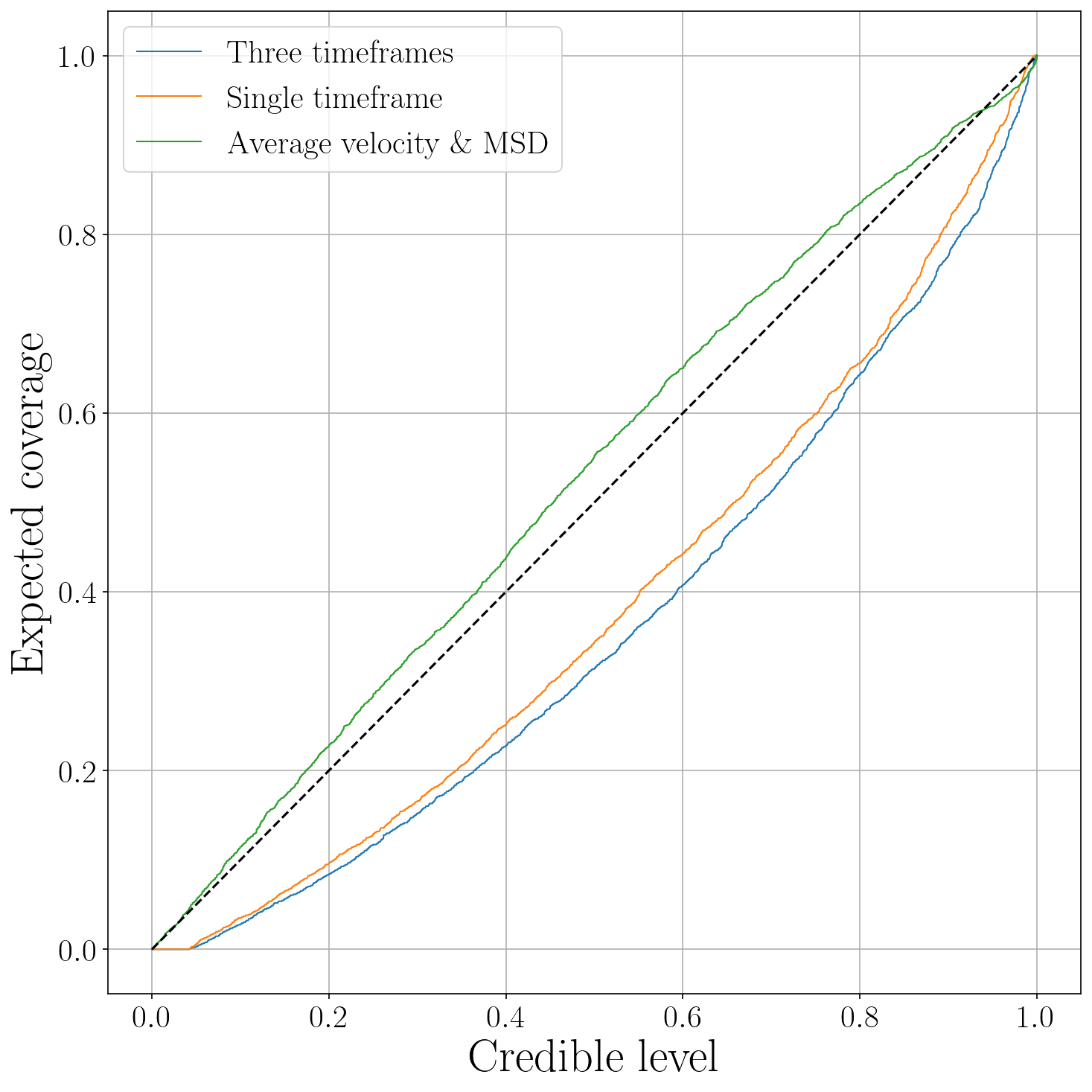}
	\end{center}
	\caption{Comparing (a) posterior pairplots for SBI using summary statistics, final interaction graph, and three snapshots of the interaction graph, and (b) a comparison of coverage tests for the different summarising methods.}
	\label{fig:3}
\end{figure}

We find that using the interaction information of particles, described by an interaction graph, is sufficient to infer parameters for active matter models. In \autoref{fig:3}, we show the the estimated posterior using three different summarising approaches for the simulation output, where we take the simulation output shown in \autoref{fig:active_matter_model}(b) and \autoref{fig:2} as ground truth. First, we find that we are able to correctly identify the true parameters using the average velocity and MSD as summary statistics but with high uncertainty, as shown in green where the probability density is spread across, for example, the prior distribution of interaction stiffness values. When we use only the last time step, shown in orange, the probability mass becomes much more centered around the true parameter values, marked as dashed lines and black dots. These results demonstrate that even a single snapshot of interactions can outperform classical methods for summarising small systems of interacting particles. 

As we use more timesteps as summarising features, our posterior estimation becomes increasingly tighter around the true parameter values, as shown as blue curves in \autoref{fig:3}. Here, we use only three timesteps, the initial structure of the system, the final structure of the system (as in orange), and halfway between these two timesteps. The overall increase in accuracy is relatively small compared to a single timestep but considerably tighter than using average velocity and MSD. In order to assess how well-calibrated our posteriors are, we compute the expected coverage, which quantifies the probability that a set of parameters will be included in the highest density region of probability of a posterior (for further details see \cite{hermans2021averting}). This is shown in \autoref{fig:3}.b. We observe that, while the estimated posterior calculated using summary statistics (average velocity and MSD) has larger spread of uncertainty, the expected coverage  is above the diagonal implying that the posterior distributions are over-dispersed. On the other hand, the expected coverage for posteriors calculated using the interaction graph are lower, indicating they are under-dispersed. We see relatively little difference in the impact of the number of snapshots used. Future work will investigate how to improve the expected coverage of these graph-informed posteriors.

\section{Discussion}
\label{discussion}

We have demonstrated that using relational data in place of summary statistics provides a method for inferring parameters of active matter. Active matter systems, ubiquitous in biology, are especially difficult to couple to mathematical models as they generate high-dimensional stochastic trajectories. These systems can exhibit a range of rheologies, from fluid to crystalline, which are thought to be important for many biological processes including during embryonic development and cancer progression. Quantifying active matter systems requires simulating large systems over long timescales and comparing point summaries of system statistics to observations.

Here, we demonstrate that geometric deep learning, combined with simulation-based inference, which we refer to as graph-informed SBI, allows for the estimation of the posterior distribution for small active matter systems over short timescales. This is the regime where the descriptive power of classical methods for summarising these systems, such as through calculating intermediate scattering functions or velocity correlations, are restricted by the small size of the system. Surprisingly, we find that even the relational data contained in the final timestep of simulation output has sufficient descriptive power to estimate parameter values. This finding is empirically supported by other work, for example in \citep{bapst2020unveiling}, which linked structure to dynamics in glassy systems using GNNs. Importantly, our work has significant value for the field of cell biology, as our results demonstrate that \textit{in vitro} experiments can be used to determine rheological impacts of changes in density even when the number of cells observed is small.

Finally, we note that our work is an exciting yet preliminary step on the path towards combining geometric deep learning with collective interacting systems. In this work, we use a GCN and embed temporal information through a spatio-temporal adjacency matrix. We expect further improvements of our method through the introduction of concepts such as attention or memory. We look forward to work in this progression and applying our approach to real experiment data.

\section{Acknowledgements}
NRS is thankful for the Alan Turing Postdoctoral Enrichment Award (184837) for funding this work. Work in RM laboratory is supported by grants from the Medical Research Council (MR/S007792/1), Biotechnology and Biological Sciences Research Council (M008517, BB/T013044) and Wellcome Trust (102489/Z/13/Z). 
 
\bibliography{active_matter_sbi}
\bibliographystyle{iclr2023_conference}

\appendix
\section{Appendix}
\label{appendix}

\subsection{Active Matter Model}\label{ap:model}

In this work, we use a custom-built simulator for active matter simulations in 2D. Taking motivation from collective cell migration, we assume that motion is over-damped. For each timestep, we update the position, according to
\begin{equation}\label{eq:langevin}
	\dot{\mathbf{r}}_i = v_0 \hat{\mathbf{n}}_i + \frac{1}{\zeta} \sum_{j} \mathbf{F}_{ij}
\end{equation}	
where $\mathbf{r}$ is the position of each particle, $v_0$ is the active force, $\mathbf{n}$ is the normal to the particle, $\zeta$ is the friction in the system, and where $\mathbf{F}_{ij}$ is the interaction potential between particles $i$ and $j$. We also update an angular variable, $\theta^r$, which is described by 
\begin{equation}
	 \dot{\theta}^r_i = \eta_i, \quad  \langle \eta_i \rangle =0,\: \: \langle \eta_i(t) \eta_j(t') \rangle = \frac{1}{\tau} \delta_{ij} \delta(t-t').
 \end{equation}
This variable causes the particle to perform a random walk with diffusion constant $D_r = \frac{1}{2\tau}$, where $\tau$ is the persistence timescale and is one of the parameters that we investigate. The noise, $\eta$ is assumed to be Gaussian and where $\delta$ is the Kronecker delta.

At each timestep, we also update the position based on distance-based interactions between other particles. The interaction forces are described by a linear piecewise force law which decreases linearly as a function of distance. We refer to the slope of decrease as the stiffness of interactions and denote this by $k$. This is also one of the parameters that we investigate, along with the persistence timescale, $\tau$, and the active propulsion force of the particle, $v_0$. For two particles $i$ and $j$ separated by a distance $r = |\mathbf{r}_j - \mathbf{r}_i|$ and with radii $R_i$ and $R_j$, where we denote the sum of radii with $b_{ij}$, we calculate the forces using the following interaction force law,

\begin{equation}
	F_{ij}(r) =
	 \begin{cases}
		k \left(r - b_{ij}\right)  & \text{if } \frac{r}{R_i +R_j} < 1+\epsilon \\
		- k \left(r - b_{ij} - 2 \epsilon b_{ij}\right) &  \text{if } 1+\epsilon < \frac{r}{R_i +R_j} < 1+2  \\
		0 & \text{otherwise}\\
	\end{cases}
	\end{equation}
Here, $\epsilon$ is a dimensionless parameter that reflects the attraction strength of interactions or the adhesion. This interaction force is calculated for particles within a contact radius given by their radius and a cutoff region equal to $1 + 2\eps$

In this work, we take cell migration as inspiration. Our interaction dynamics is inspired by biological cells, as in \citep{matoz2017cell}, and we dimensionalise simulation units to be appropriate for modelling cell assays. Hence, cell radius is 10 microns and the velocity of cells is in units microns/hour. To generate simulation output, we sample from a uniform prior distribution where both $v0$ and $k$ are between 0 and 150 and $\tau$ is between 0 and 10. we construct the system such that particles have polydispersity of 0.3. Finally, we assume that the number of particles remains unchanged and that the system domain is open.\\

\subsection{Neural Posterior Architectures}\label{ap:architectures}

We consider three different inference approaches to estimate the posterior distribution of our. For all three, we use neural posterior estimation (NPE) with a masked autoregressive flow (MAF) \citep{papamakarios2017masked, greenberg2019automatic}. For all architectures, we pass the NPE network observational features of size 100. Where we use only summary statistics, this is the average velocity and mean square displacement and for graph-informed statistics, we embed the interaction networks into graph-embeddings of size 100. 

We first consider the influence of architecture size for the inference of the posterior using only summary statistics. We find that we get best performance, as measured by the negative log-likelihood for MAFs with four layers of size 256, where we also consider fewer (3) and smaller (64,128) layers. Having found the best performance for summary statistics, we fix the inference network shape to allow for fair comparison against embedding architectures.

For the embedding network, we use a graph-convolutional network \citep{kipf2016semi} with 3 steps of message passing and with batch normalisation before each step. We also compute the unified graph embeddings as a final step using global (mean) pooling. As we discuss in \autoref{discussion}, further work will explore different architectures. Here, we alter only the relative size for each message passing layer (between 64,128 and 256). We find that large networks (256) give the best performance.

\begin{figure}[h]
	\begin{center}
		\includegraphics[width=0.75\linewidth]{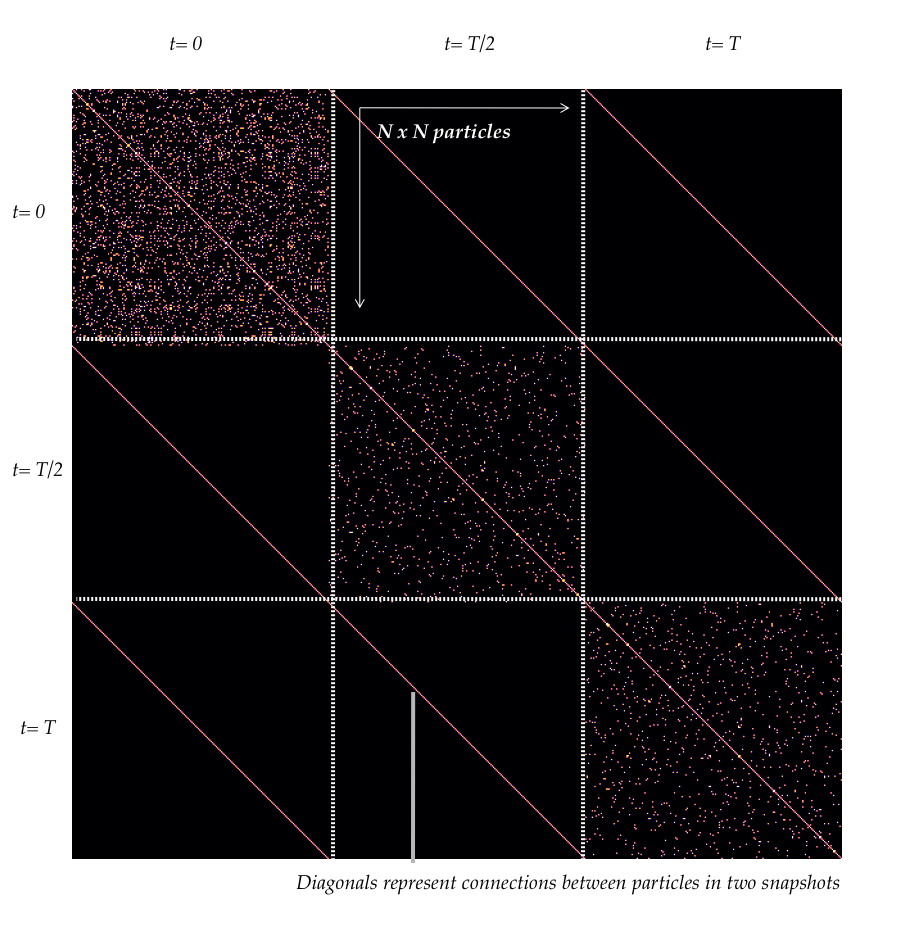}		
	\end{center}
	\caption{Annotated spatio-temporal adjacency matrix where each diagonal block represents an interaction network for single time snapshot and where off diagonals represent tracked particle positions between $t$ to $t+1$. }
	\label{fig:4}
\end{figure}

Finally, we pass the graph embedding networks the position and velocity for all particles. In this work, temporal interaction data is encoded in the adjacency matrix. We construct block diagonal matrices, where each block reflects the spatial interaction network between particles $i$ and $j$ at a single timestep. The off-diagonals are also block matrices but connect individual particles between timesteps $t$ and $t+1$. An annotated spatio-temporal adjacency matrix is shown in \autoref{fig:4}. 

\end{document}